\def\etal{et~al.\ }
\def\CIVdblt{{\rm C}\kern 0.1em{\sc iv}~$\lambda\lambda 1548, 1550$}
\def\MgIIdblt{{\rm Mg}\kern 0.1em{\sc ii}~$\lambda\lambda 2976, 2803$}
\def\NVdblt{{\rm N}\kern 0.1em{\sc v}~$\lambda\lambda 1238, 1242$}  
\def\OVIdblt{{\rm O}\kern 0.1em{\sc vi}~$\lambda\lambda 1031, 1037$} 
\def\SiIVdblt{{\rm Si}\kern 0.1em{\sc iv}~$\lambda\lambda1393, 1402$}  
\def\AlI{\hbox{{\rm Al}\kern 0.1em{\sc i}}}
\def\AlII{\hbox{{\rm Al}\kern 0.1em{\sc ii}}}
\def\AlIII{{\hbox{\rm Al}\kern 0.1em{\sc iii}}}
\def\CaII{\hbox{{\rm Ca}\kern 0.1em{\sc ii}}}
\def\CII{\hbox{{\rm C}\kern 0.1em{\sc ii}}}
\def\CIIe{\hbox{{\rm C$^{\ast}$}\kern 0.1em{\sc ii}}}
\def\CIII{\hbox{{\rm C}\kern 0.1em{\sc iii}}}
\def\CIV{\hbox{{\rm C}\kern 0.1em{\sc iv}}}
\def\CV{\hbox{{\rm C}\kern 0.1em{\sc v}}}
\def\NHI{\hbox{$N$({\rm H}\kern 0.1em{\sc i})}} 
\def\HI{\hbox{{\rm H}\kern 0.1em{\sc i}}}
\def\HII{\hbox{{\rm H}\kern 0.1em{\sc ii}}}
\def\Lya{\hbox{{\rm Ly}\kern 0.1em$\alpha$}}
\def\Lyb{\hbox{{\rm Ly}\kern 0.1em$\beta$}}
\def\Lyg{\hbox{{\rm Ly}\kern 0.1em$\gamma$}}
\def\Lyfive{\hbox{{\rm Ly}\kern 0.1em$5$}}
\def\Lysix{\hbox{{\rm Ly}\kern 0.1em$6$}}
\def\Lyseven{\hbox{{\rm Ly}\kern 0.1em$7$}}
\def\Lyeight{\hbox{{\rm Ly}\kern 0.1em$8$}}
\def\Lynine{\hbox{{\rm Ly}\kern 0.1em$9$}}
\def\Lyten{\hbox{{\rm Ly}\kern 0.1em$10$}}
\def\HeI{\hbox{{\rm He}\kern 0.1em{\sc i}}}
\def\HeII{\hbox{{\rm He}\kern 0.1em{\sc ii}}}
\def\FeI{\hbox{{\rm Fe}\kern 0.1em{\sc i}}}
\def\FeII{\hbox{{\rm Fe}\kern 0.1em{\sc ii}}}
\def\FeIII{\hbox{{\rm Fe}\kern 0.1em{\sc iii}}}
\def\MnII{\hbox{{\rm Mn}\kern 0.1em{\sc ii}}}
\def\MgI{\hbox{{\rm Mg}\kern 0.1em{\sc i}}}
\def\MgII{\hbox{{\rm Mg}\kern 0.1em{\sc ii}}}
\def\MgIII{\hbox{{\rm Mg}\kern 0.1em{\sc iii}}}
\def\NV{\hbox{{\rm N}\kern 0.1em{\sc v}}}
\def\NII{\hbox{{\rm N}\kern 0.1em{\sc ii}}}
\def\NIII{\hbox{{\rm N}\kern 0.1em{\sc iii}}}
\def\OVI{\hbox{{\rm O}\kern 0.1em{\sc vi}}}
\def\OII{\hbox{[{\rm O}\kern 0.1em{\sc ii}]}}
\def\SiI{\hbox{{\rm Si}\kern 0.1em{\sc i}}}
\def\SiII{\hbox{{\rm Si}\kern 0.1em{\sc ii}}}
\def\SiIII{\hbox{{\rm Si}\kern 0.1em{\sc iii}}}
\def\SiIV{\hbox{{\rm Si}\kern 0.1em{\sc iv}}}
\def\SII{\hbox{{\rm S}\kern 0.1em{\sc ii}}}
\def\SIII{\hbox{{\rm S}\kern 0.1em{\sc iii}}}
\def\SIV{\hbox{{\rm S}\kern 0.1em{\sc iv}}}
\def\NaI{\hbox{{\rm Na}\kern 0.1em{\sc i}}}
\def\kms{\hbox{km~s$^{-1}$}}
\def\cm2{\hbox{cm$^{-2}$}}

\def\ga{\mathrel{\raise.3ex\hbox{$>$}\mkern-14mu
             \lower0.6ex\hbox{$\sim$}}}

\def\la{\mathrel{\raise.3ex\hbox{$<$}\mkern-14mu
             \lower0.6ex\hbox{$\sim$}}}     
\documentstyle[11pt,aaspp4]{article}

\tighten
\begin{document}



\lefthead{Ganguly, Churchill \& Charlton}
\righthead{An Aluminum Enhanced Cloud}


\title{An Aluminum Enhanced Cloud in a {\CIV} Absorber
at $z = 1.94$\altaffilmark{1}}

\thispagestyle{empty}

\author{Rajib~Ganguly, 
        Christopher~W.~Churchill\altaffilmark{2}, 
        and 
        Jane~C.~Charlton}

\affil{Astronomy and Astrophysics Department \\
       Pennsylvania State University,
       University Park, PA 16802 \\
       {\it ganguly, cwc, charlton@astro.psu.edu}}

\altaffiltext{1}{Based in part on observations obtained at the
W.~M. Keck Observatory, which is jointly operated by the University of
California and the California Institute of Technology.}
\altaffiltext{2}{Visiting Astronomer, The W.~M.~Keck Observatory}


\begin{abstract}

In the $z=1.94$ {\CIV} absorption line system in the spectrum of quasar
Q1222+228 ($z_{em}=2.04$), we find two clouds which have contrasting physical
conditions, although they are only at a 17~{\kms} velocity separation. In the
first cloud {\SiII}, {\SiIV}, and {\CII} are detected, and {\AlII} and {\AlIII}
column density limits in conjunction with photoionization models allow us to
infer that this cloud has a large Si abundance and a small Al abundance
relative to a solar abundance pattern. This pattern resembles that of
Galactic metal--poor halo stars, which must have formed from such high redshift
gas. The second cloud, in contrast, has detected {\AlII} and {\AlIII} (also
{\SiIV} and {\CII}), but no detected {\SiII}. We demonstrate, using
photoionization models, that Al/Si must be greater than (Al/Si)$_\odot$ in
this unusual cloud. Such a ratio is not found in absorption profiles looking
through Milky Way gas. It cannot be explained by dust depletion since Al
depletes more severely than Si. Comparing to other Al--rich environments, we
speculate about the processes and conditions that could give rise to this
abundance pattern.
\end{abstract}



\section{Introduction}
\label{sec:intro}

One of the motives for studying quasar (QSO) absorption line systems is to
document the photoionization structure, chemical composition, and kinematics of
the ISM and halos of high redshift galaxies at a level of detail on par with
studies of the Milky Way ISM and Halo. QSO lines of sight that pass through
high redshift galaxies sample multiple gaseous structures having a variety of
physical conditions such as found in Galactic {\HI} and {\HII} regions,
supershells, infalling Halo gas, and material being processed at the
Galaxy/Halo interface. Recognizing the abundance patterns, photoionization
conditions, and kinematics associated with these various types of structures is
a prerequisite for a detailed understanding of the evolution of galactic gas.

Based upon low resolution spectra, early QSO absorption line efforts have been
limited to simple curve of growth analyses of equivalent widths, providing only
a weighted--average of the absorbing gas properties in each system.
A great deal has been learned from these studies and a global (statistically
based) picture of chemical and ionization evolution has been suggested [see
Steidel (1993\nocite{steidel93}) and references therein]. The next logical step
is to examine {\it variations\/} in the chemical and ionization conditions
within single galaxies and to incorporate kinematic information. Using high
resolution UV spectra, several researchers studied the Milky Way Disk and Halo
at a detailed level (\cite{wel97}; \cite{savaraa}; \cite{snow96}; \cite{sf95};
\cite{fs94}; \cite{sf93}). Some recent efforts have focused on the
cloud--by--cloud conditions in high redshift galaxies (cf.~\cite{tripp97};
\cite{pet94}). It is hoped that, ultimately, a statistical picture of the
processes that give rise to the observed absorbing gas properties will improve
our understanding of the present--epoch Milky Way and galactic evolution. In
this {\it Letter\/} we study the cloud--to--cloud properties in a {\CIV} system
at $z \sim 1.94$ along the line of sight toward the quasar Q$1222+228$ in order
to: 1) demonstrate a large variation in ionization and/or abundance conditions
between two kinematically adjacent clouds in the same absorber; and 2) present
an unusual cloud that has an Al/Si abundance ratio enhanced by a factor of
several relative to the solar ratio.

\section{The Data}
\label{sec:data}

Our spectrum of Q$1222+228$, which has resolution $\Delta v _{\rm FWHM} = 6.6$~
{\kms} with three pixels per resolution element, is a 3600 second exposure
obtained with HIRES/Keck I ({\cite{vog94}) on 23 January 1995. The wavelength
coverage is 3810.5--6304.9~{\AA} with breaks redward of 5100~{\AA}. The
spectrum was obtained as part of a large study of intermediate redshift {\MgII}
absorbers (\cite{cwcthesis}), and thus the captured wavelength coverage did not
include certain desirable transitions of the $z \sim 1.94$ {\CIV} system. The
reduction and analysis of the data are described elsewhere (Churchill
1997\nocite{cwcthesis}, Churchill, Vogt, \& Charlton 1998a\nocite{cvc98},
Schneider \etal 1993, Churchill \etal 1998b\nocite{crcv98}). In
Figure~\ref{fig:thedata}, we present the detected transitions (5$\sigma$)
aligned in line--of--sight velocity.

We focus upon the {\SiII} $\lambda 1527$, {\AlII} $\lambda 1671$, and {\AlIII}
$\lambda 1863$ transitions in two absorbing clouds centered at $v\sim 0$
and $\sim 17$~{\kms}. The {\CIVdblt} profiles are saturated across this
velocity interval, but the {\SiIV} and {\CII} profiles clearly show two
distinct clouds. In Cloud A, at $v\sim 0$~{\kms}, we have detected
{\SiII}~$\lambda 1527$ and obtained upper limits on both {\AlII} $\lambda 1671$
and {\AlIII} $\lambda 1863$. In Cloud B, at $v \sim 17$~{\kms}, the converse is
true; we have detected {\AlII} $\lambda 1671$ and {\AlIII} $\lambda 1863$,
and obtained an upper limit on {\SiII} $\lambda 1527$. The
{\AlIII} $\lambda 1855$ transition, the doublet counterpart of
{\AlIII} $\lambda 1863$, was not covered.

The cloud velocities, column densities, and Doppler $b$ parameters were
obtained from Voigt profile (VP) decomposition of the spectra using the program
MINFIT (\cite{cwcthesis}), which minimizes $\chi^2$ between the model and the
data. All detected transitions were fit simultaneously, with the exception of
the {\CIV} $\lambda 1550$ transition, which is blended with an unidentified
feature. The VP results are listed in Table~\ref{tab:vpfits}. The $5\sigma$
equivalent width limits for the non--detected transitions were converted to
column density upper limits using the curve of growth. Since these
equivalent width limits correspond to the linear part of the curve of growth,
the column density limits are not sensitive to the adopted $b$ parameter
(see Table~\ref{tab:vpfits}). We note that $N({\SiII})$ also corresponds to the
linear part of the curve of growth and therefore is well constrained despite
the large fractional error in $b({\SiII})$.

Constraints are available for the total neutral hydrogen column density of the
absorber, which can be interpreted as an upper limit on $N({\HI})$ of any
single cloud in the system. 
In the FOS/HST spectrum of Impey \etal (1996\nocite{impey96}), there
is no indication of a break at the expected location of the Lyman
limit. 
They report detections of {\Lyb} and {\Lyg} that are members of
complex blends which suffer from a pre--COSTAR point spread
function. 
Fern\'{a}ndez--Soto \etal (1995\nocite{alberto}) performed a profile
fit to a saturated, multi--component {\Lya} feature at $\sim 3571$~{\AA} in
a spectrum with resolution $\sim 30$~{\kms}.
However, the number of components and their $b$ parameters and column
densities are too uncertain to provide a constraint more useful than
that provided by the lack of a Lyman limit break.
We have conservatively adopted an upper limit of $\log N({\HI})
\sim 16.5$~{\cm2}.


\section{Photoionization Modeling of the Clouds}
\label{sec:models}

We inferred the clouds' physical conditions using CLOUDY (\cite{ferland}).  
We have assumed a $z=2$~extragalactic UV ionizing background, given by Haardt
\& Madau (1996\nocite{handm96}) ($\log F_{\nu} = -20.35$ erg~s$^{-1}$~Hz$^{-1}$
~cm$^{-2}$). The input parameters for CLOUDY are the neutral hydrogen
column density, $N({\HI})$, the metallicity, $Z$, and the total number density
of hydrogen, $n_{H}$. CLOUDY assumes that photons are incident on one side
of a plane--parallel slab of gas with constant $n_{H}$ and integrates the
equations of radiative transfer along an optical path through the slab until it
reaches a specified $N({\HI})$. For a fixed spectrum, there is a relationship
between $n_{H}$ and the ionization parameter $U$ (the number density of
hydrogen ionizing photons over the number density of electrons).
For simplicity, we adjusted the two ratios\footnote{We use
the notation $\hbox{[X/H]} = \log (\hbox{X/H}) - \log (\hbox{X/H})
_{\odot}$.} [Si/H] and [Al/H] keeping $\hbox{[C/H]}=0$. An important
assumption is that the modelled transitions arise in a single phase
environment. The broad, saturated, {\CIV} profile may arise in a physically
distinct higher ionization phase from the {\CII} and other lower ionization
species. We argue that the precise velocity alignment of the lower ionization
species and the fact that their $b$ parameters are consistent (see
Table~\ref{tab:vpfits}) justify the assumption.


For Cloud A, we measured $N({\SiII})$, $N({\SiIV})$, $N({\CII})$ and upper
limits on $N({\AlII})$ and $N({\AlIII})$. Without precise information on the
neutral hydrogen {\it distribution\/} in the system, we modelled the cloud for
a range of $N({\HI})$. For a given $N({\HI})$ we ran CLOUDY in its
``optimized'' mode, allowing $Z$, $n_{H}$, and [Si/H] to vary until the cloud
model was consistent with the observed $N({\SiII})$, $N({\SiIV})$, and
$N({\CII})$. The ratio $N({\SiII})/N({\SiIV})$ constrained both the ionization
parameter and metallicity. For models with $\log N({\HI}) > 15.4$~{\cm2},
$N({\AlII})$ was greater than the upper limit. We therefore set all other
parameters equal to those given by the optimized solution and then solved for
the model cloud by varying [Al/H] and optimizing on the measured upper limits
on $N({\AlII})$ and $N({\AlIII})$. In the upper panel of
Figure~\ref{fig:model}, we illustrate how the [Si/H] and [Al/H] ratios, $U$,
and $Z$, must be scaled for the models to be consistent with the data. For
$\log N({\HI}) \geq 15.5$~{\cm2}, a factor of two to three enhancement of
[Si/H] and an underabundance of [Al/H] are obtained, nearly independent of
metallicity. The lack of a Lyman limit break in the FOS/HST spectrum of Impey
\etal (1996\nocite{impey96}) precludes a large $N({\HI})$. This cloud could be
consistent with [Si/H] and [Al/H]$\sim 0$ (solar values) only if the
metallicity is super--solar [for $\log N({\HI}) \sim 15$~{\cm2}]. We conclude
that [Si/H] is enhanced and [Al/H] is underabundant by a factor of a few for
the more plausible range of metallicities, $-1.4 \leq \hbox{[Z/Z$_{\odot}$]}
\leq -0.3$. This pattern is similar to that seen in Galactic Halo stars of
comparable metallicities (\cite{lau96}; \cite{savaraa}).


For Cloud B, we have measured $N({\AlII})$, $N({\AlIII})$, $N({\CII})$,
$N({\SiIV})$ and an upper limit on $N({\SiII})$. As with Cloud A, the Lyman
limit break provides the constraint of $\log N({\HI}) \leq 16.5$~{\cm2}.
Al is unusual in that its high--temperature dielectronic recombination rates
are high and the ratio $N({\AlII})/N({\AlIII})$ can increase with increasing
$U$~for $\log N({\HI}) \geq 17$~{\cm2} (\cite{pet94}). For smaller $N({\HI})$
this is not the case; a decreasing $U$~yields an increasing $N({\AlII})$ as
naively expected. Nonetheless, because of the unusual nature of reactions
involving Al, we approached the modeling of Cloud B with caution.

First, we considered the constraints that can be placed on the model cloud
without considering the Al column densities (i.e.\ using only the {\CII} and
{\SiIV} detections and the {\SiII} limit, we allowed $Z$, $n_H$, and [Si/H] to
vary). We found models consistent with the data for $15.5 \leq \log N({\HI})
\leq 16.75$~{\cm2}.  Smaller $N({\HI})$ values were unable to produce enough
{\SiIV} relative to {\CII}. For larger $N({\HI})$ values, $U$~must be larger,
corresponding to a smaller $n_H$, and the model cloud was unrealistically large
and Jeans unstable. For the acceptable range of $N({\HI})$, we found
$ -0.3 \leq \hbox{[Si/H]} \leq -0.5$. The metallicity ranged from
$\hbox{[Z/Z$_{\odot}$]} = +0.2$ for $N({\HI}) = 15.5$~{\cm2} to
$\hbox{[Z/Z$_{\odot}$]} = -1.5$ for $N({\HI}) = 16.75$~{\cm2}. Next, we
considered what model clouds were consistent with the constraints provided by
only the {\AlII} and {\AlIII} column densities while allowing $Z$ and $n_H$ to
vary. Again, a consistent model cloud could be achieved for $15.0 \leq \log
N({\HI}) \leq 16.75$~{\cm2}. Systematically, the required metallicities were
half to one dex larger than those required by the C and Si constraints alone.
Naively, these two experiments imply [Al/Si]$>0$, consistent with the striking
strength of the Al profiles for Cloud B.

When we ran CLOUDY in optimize mode, using all Si, C, and Al constraints and
allowing $Z$, $n_H$, [Si/H], and [Al/H] to vary, we found $\hbox{[Si/H]} < 0$
and $\hbox{[Al/H]} > 0$ were required for all possible $N({\HI})$. Only the
narrow range $15.9 \leq \log N({\HI}) \leq 16.25$~{\cm2} yielded acceptable
cloud models. For smaller values of $N({\HI})$, the ratios
$N({\SiIV})/N({\SiII})$ and $N({\AlIII})/N({\AlII})$ could not be made
consistent with the data (for the same $U$). The model results are illustrated
in the lower panel of Figure~\ref{fig:model}. Over the permitted range of
$N({\HI})$, we have $-0.55 \leq \hbox{[Z/Z$_{\odot}$]} \leq -0.32$,
$-2.5 \leq \log U \leq -2$, $\hbox{[Al/H]} \sim +0.3$, and
$-0.6 \leq \hbox{[Si/H]} \leq -0.2$.

Given the unusual conclusion we have reached that the Al in Cloud B is enhanced
by a factor of two relative to solar, we now consider the validity of our
assumption of photoionization equilibrium. Trapero \etal (1996\nocite{trap96})
found the ratio of $N({\AlIII})/N({\AlII})$ out of equilibrium in the
interstellar cloud toward the B1~V star 23~Ori. In that case, excited
carbon implied a high density so that collisional ionization was applicable;
the ratio of $N({\AlIII})/N({\AlII})$ implied a higher temperature than that
measured from Doppler parameters, indicating that the gas had cooled faster
than it could recombine. We obtained an unrestrictive upper limit on the
density in Cloud B from an upper limit on $N({\CIIe})$. However, we have no
reason to believe Cloud B suffers from such a non--equilibrium situation.
Cloud B is consistent with photoionization in that the CLOUDY models are found
to have kinetic temperatures consistent with those inferred from the Doppler
parameters under the assumption of thermal broadening [such was not the case
for the models of Trapero \etal (1996\nocite{trap96})].

Since the clouds are likely to be in a galaxy, it is possible that stellar
sources could affect the shape of the ionizing radiation spectrum. However, the
following basic argument applies. Both {\AlI} and {\SiI} have similar
ionization potentials (6.0~eV and 8.2~eV, respectively) as do {\AlII} and
{\SiII} (18.8~eV and 16.3~eV, respectively). Unless the spectrum of ionizing
radiation has a very abrupt feature in these energy ranges, {\AlII} and
{\SiII} will be similarly affected by a change in spectral shape. We ran test
models and confirmed this with a star--forming galaxy spectrum
(\cite{bruzual}).


\section{Summary and Speculations}
\label{sec:summary}

Two kinematically adjacent, absorbing clouds in the same galaxy at
$v\sim0$~{\kms} and $\sim 17$~{\kms} have very different abundance patterns,
if they are in photoionization equilibrium. Based upon CLOUDY models, Cloud A
is inferred to have Si enhanced relative to a solar abundance pattern, and Al
underabundant relative to solar. This abundance pattern is characteristic of
Milky Way Halo stars (\cite{lau96}; \cite{savaraa}) which could have formed
from gas like that observed in this $z=1.94$ galaxy. We would expect such a
pattern for many clouds in high redshift galaxies. Independent of the details,
Cloud B has Al several times enhanced relative to Si (compared to the solar
abundance pattern). This Al enhancement is highly unusual. If depletion onto
dust grains is important we would expect a smaller Al/Si since Al depletes more
readily than Si. To date, such an Al enhanced pattern has not been seen in
absorption profiles looking through Milky Way interstellar gas (\cite{savaraa};
\cite{snow96}; \cite{sf95}; \cite{fs94}; \cite{sf93}).

How could an enhancement of Al originate in a $z=1.94$ cloud? In an attempt to
find clues we note three astrophysical environments in which Al enhancement is
observed: 1) in the stellar photospheres of the most metal poor globular
clusters (\cite{shet96}; \cite{smith96}); 2) in the broad line regions of some
AGNs\footnote{We note that the absorber lies 30,000 {\kms} from the emission
redshift of the QSO, which is not a BAL AGN.} (\cite{shields97}); and 3) in the
photospheres of Milky Way Bulge stars (\cite{mr94}). The common theme for the
enhancement of Al in these three seemingly different environments is novae. The
novae could produce the Al either directly in their ejecta (\cite{smith96}) or
indirectly by providing magnesium isotopes to the ISM that later deposit onto
stellar photospheres (\cite{langer}). The $^{26}$Mg and $^{27}$Mg
isotopes would then be converted to Al through proton capture in deep CNO
convective mixing layers in metal poor stars (\cite{langer}). More generally,
does this suggest that a concentration of novae contributed to enhancing the Al
in this cloud? In the globular cluster environment one key is to have a large
number of stars that were formed coevally. Another key, which may apply to all
three environments, is to have a potential well large enough to retain the gas.
These may be prerequisites in our case also. Another possibility for excess Al
production is a particular class of supernova for progenitors over a narrow
mass range. The predicted amount of enhancement relative to other elements
depends on the specific supernovae model adopted (\cite{nom97}), however only
a small subclass of models would give Al/Si in agreement with that inferred for
this cloud.

Just how unusual is this cloud with a large Al/Si ratio?  
Such a pattern has not been seen in absorption along dozens of lines
of sight toward Milky Way disk and halo stars.
Most high redshift clouds do not have large {\AlIII} and {\AlII}
(relative to {\SiII}).
However, in the $z = 2.14$ damped {\Lya} absorber in Q$0528-251$
(\cite{lu96}), we have identified a single outlying cloud (at $\sim
-100$~{\kms}) for which the equivalent width of the {\AlII} $\lambda
1671$ transition is larger than that of the {\SiII} $\lambda 1527$
transition.
The other clouds in the same system clearly have the opposite ratio.
Finding more examples and establishing similarities between Al--rich
environments are logical next steps in diagnosing the origin of this
abundance pattern at high redshift and perhaps understanding the
anomalous enhancements seen in metal--poor globular cluster and
Bulge stars.


\acknowledgments

This work was supported by NSF AST--9529242 and AST--9617185 and by NASA
NAG5--6399.  Thanks to U.~Hellsten for providing the Haardt \& Madau input
spectrum, and to  A. Fern\'{a}ndez--Soto, J. Lauroesch, M. Rich,
B. Savage, D. Schneider, S. Sigurdsson, R. Wade, D. Welty for
collectively contributing a breadth of knowledge.   Special thanks to
Steven S. Vogt for his years dedicated to building HIRES.


\onecolumn


\begin{deluxetable}{lccccc}
\tablewidth{0pc}
\tablecaption{Column Densities of Key Transitions}
\tablehead
{
\colhead{} &
\multicolumn{2}{c}{Cloud A} & &
\multicolumn{2}{c}{Cloud B} \\
\cline{2-3} \cline {5-6}
\colhead{Transition} &
\colhead{$log(N~{\rm cm}^{-2})$} &
\colhead{$b$ ({\kms})} & &
\colhead{$log(N~{\rm cm}^{-2})$} &
\colhead{$b$ ({\kms})} 
}
\startdata
{\AlII} 1670.787 & $<11.13$\tablenotemark{a} & \nodata & & $11.85 \pm 0.04$ & $4.91 \pm 0.60$ \nl 
{\AlIII} 1862.790 & $<11.79$\tablenotemark{a} & \nodata & & $12.06 \pm 0.10$ & $5.17 \pm 1.71$ \nl
{\CII}  1334.532 & $13.11 \pm 0.11$ & $4.80 \pm 2.02$ & & $13.45 \pm 0.07$ & $6.91 \pm 1.44$ \nl  
{\SiII} 1526.707 & $12.42 \pm 0.11$ & $1.84 \pm 1.88$ & & $<12.29$\tablenotemark{b} & \nodata \nl
{\SiIV} 1393.755 & $12.72 \pm 0.04$ & $4.70 \pm 0.74$ & & $13.11 \pm 0.03$ & $6.50 \pm 0.53$ \nl
{\SiIV} 1402.770  & $12.72 \pm 0.04$ & $4.70 \pm 0.74$ & & $13.11 \pm 0.03$ & $6.50 \pm 0.53$ \nl
\enddata
\tablenotetext{a}{Upper limit for assumed $b=5~{\kms}$, based upon the {\CII} $\lambda  1334$ transition}
\tablenotetext{b}{Upper limit for assumed $b=5~{\kms}$, based upon the {\AlII} $\lambda 1670$ transition}
\label{tab:vpfits}
\end{deluxetable}

\begin{figure}[th]
\plotfiddle{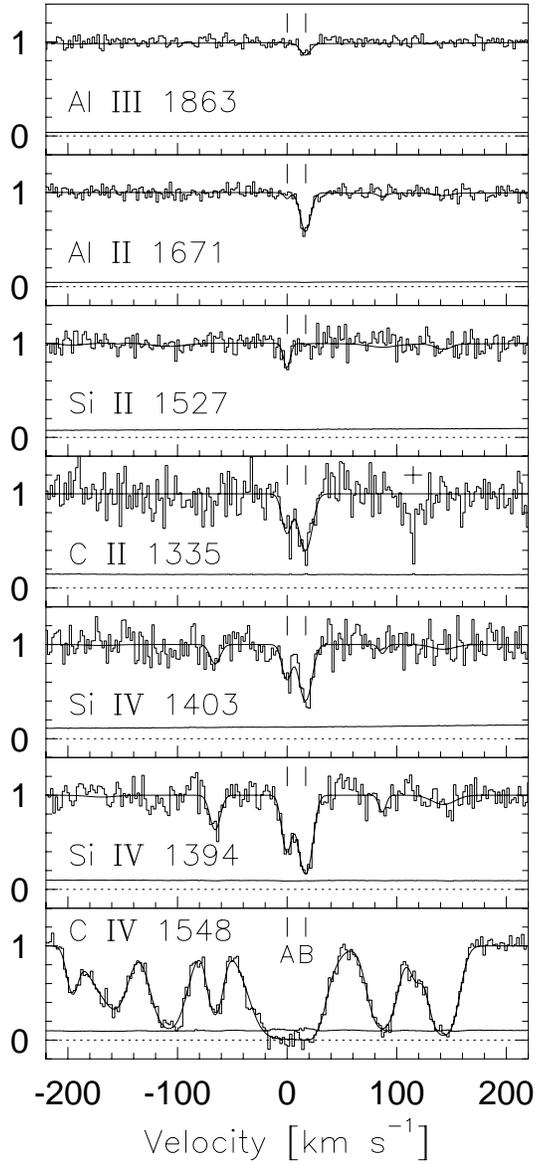}{6.0in}{0}{65.}{65.}{-200}{-27}
\protect\caption[]{The HIRES/Keck absorption profiles of the $z=1.94$
{\CIV} system in the spectrum of Q$1222+228$.  The two clouds studied
here, A and B, are marked by the vertical ticks at $v \sim
0$~{\kms} and $v \sim 17$~{\kms}, respectively.  The cross mark at
$v=115$~{\kms} near the {\CII} $\lambda 1335$ transition flags an
unidentified absorption feature.}
\label{fig:thedata}
\end{figure}

\begin{figure}[th]
\plotfiddle{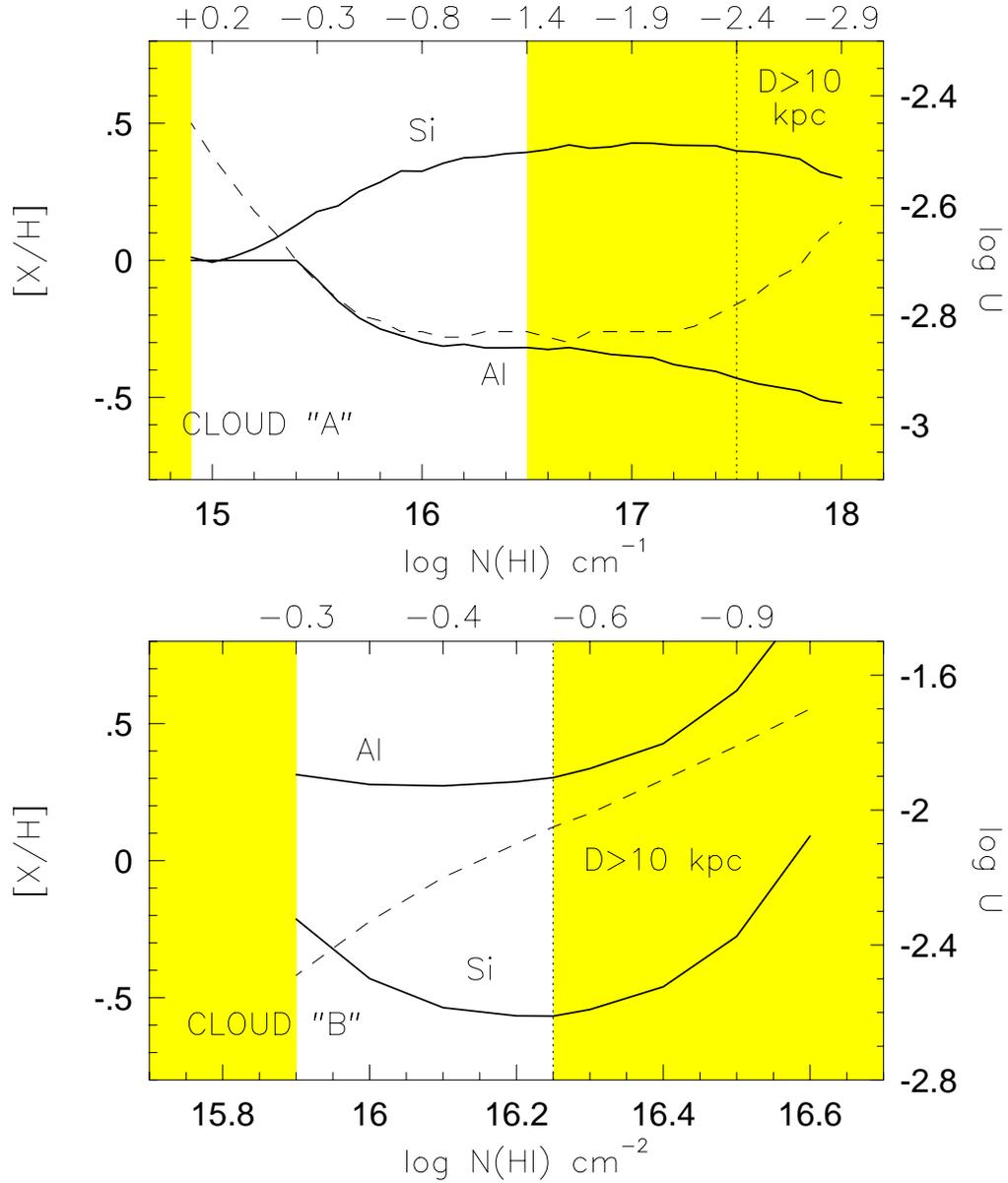}{6.0in}{0}{65.}{65.}{-200}{-27}
\protect\caption[]{The [Al/H] and [Si/H] abundance ratios (relative to
solar) as a function of $\log N({\HI})$ for clouds A (upper panel)
and B (lower panel).  The corresponding metallicities,
[$Z/Z_{\odot}$], are given by the upper horizontal axes.  The unshaded
regions give the range of models that are consistent with the observed
data (see text).  The long--dash curves are the ionization parameters,
$\log U$, which are given by the right hand vertical axes.  The
$N({\HI})$ at which the model cloud sizes exceed 10~kpc and become
Jeans unstable are marked by the vertical short--dash lines.}
\label{fig:model}
\end{figure}


\end{document}